\documentclass[twocolumn,floats,aps]{revtex4}
\usepackage{epsfig,amsmath}

\chardef\anciennecat=\catcode`\@
\catcode`\@=11
\def\section{%
  \@startsection
    {subsubsection}%
    {1}%
    {\z@}%
    {0.8cm \@plus1ex \@minus .2ex}%
    {0.5cm \@plus1ex \@minus.2ex}%
    {\normalfont\small\bfseries}}

\catcode`\@=\anciennecat

\newcommand{\as}{\alpha_s}
\newcommand{\ab}{\bar{\alpha}}
\newcommand{\G}{\Gamma}
\newcommand{\real}{{\mathcal R}{\mathrm e}}

\begin{document}

\title{\bf Forward jets in the colour-dipole model}
\author{S. Munier}
\email{munier@spht.saclay.cea.fr}
\affiliation{
 Service de Physique Th\'eorique,
 CEA/Saclay, F-91191 Gif-sur-Yvette Cedex,
 France.
}
\begin{abstract}
We show that a forward jet with large transverse momentum
in an onium-onium collision is a hard probe
which can be effectively characterized by a colour-dipole distribution
at the time of the interaction.
The dipole distribution is computed, and compared to its counterpart
for a virtual photon in the initial state.
We find that while in the photon case, the tail of large sizes is
exponentially cut-off, it contributes sizeably in the forward-jet case, 
which signs the sensitivity of observables based on such events
to the infrared region. Moreover, a direct probabilistic
interpretation of the dipole distribution fails since it takes
negative values in the large size region.
\end{abstract}

\maketitle

\section{Introduction.}

The physics at HERA has proved the successes of (resummed) perturbative QCD in describing
many observables accurately measured there: inclusive ones like the structure function
$F_2$, but also more exclusive ones, like the diffractive structure function
$F_2^D$, or heavy meson production. The justification for relying on a perturbative 
development is that the deep-inelastic scattering process naturally provides
a well-controlled
hard scale given by the photon virtuality $Q^2$, which makes the
effective strong coupling constant $\alpha_s(Q^2)$ small enough.
On the one hand, the physics of these observables 
is usually well described by the renormalization group
evolution \cite{Gribov:1972ri,Dokshitzer:1977sg,Altarelli:1977zs}
between a lower scale $Q_0^2$ at which
the proton parton densities are parametrized and the scale $Q^2$. 
On the other hand, the cross-section for
the events selected with the requirement that
a forward jet of transverse momentum
${\vec q}^2$ of the order of $Q^2$ be present in the final state is 
seemingly not described by a
straightforward DGLAP evolution 
(see ref.\cite{Potter:1999kt} and references therein): 
as a matter of fact, these Regge-like kinematics are expected to select the
BFKL dynamics \cite{Kuraev:1976ge,Kuraev:1977fs,Balitsky:1978ic}.

In $p-\bar{p}$ collisions at the Tevatron, no hard scale is provided
by the initial state. However, it can be generated in the scattering 
and manifests itself
in the final state in the form of a jet with a large transverse momentum. 
Events of this class are also accessible to
a perturbative QCD interpretation.

In this letter, we focus on high-energy onia (massive $q\bar q$
states) collisions, as a model
for $p-\bar{p}$ collisions at high energy when high-mass scales
are selected by forward jets.
The inclusive cross-sections for onia collisions have been described
using a dipole cascade modelling the rapidity evolution of the 
$q\bar q$ pairs before
their interaction \cite{Mueller:1994rr}. Here we require that
at least the first (most forward) gluon which goes to the final 
state has its transverse momentum larger than a scale $\mu$; this 
gluon becomes a forward-jet, and we interpret it as an effective 
colour-dipole distribution present at the time of the scattering.

In section {\bf 2}, we detail the modelisation that we adopt for 
the forward jet. We show in section {\bf 3} how we can extract 
to double-leading logarithmic (DLL) approximation, the dipole 
content of such an object.
Section {\bf 4} contains our conclusions and outlook.

\section{Emission of a forward gluon in the final state.}

The onium-onium forward scattering amplitude involves the exchange of
two gluons between the initial onia. At high-energy, one has to
take into account the possibility of multiple splittings of these
$t$-channel gluons. This can be done either by ($k_\bot$)factorizing 
\cite{Catani:1991eg} a
BFKL-like ladder between the bare onia, or equivalently by
computing the two-gluon exchange diagram between the onia
dressed by an arbitrary number of soft ``sea'' gluons. The latter
approach inspired the colour-dipole model of
ref.\cite{Mueller:1994rr}. 
The equivalence between these two methods was shown on different 
features of both pictures in 
ref.\cite{Chen:1995pa,Navelet:1997xn,Navelet:1997tx,Munier:1998vk}.

In this section, we shall derive in a $t$-channel picture the gluon
density $f$ which is to be considered in an interacting 
``onium+forward-jet'' system as the starting-point of a BFKL evolution. 
The next section will be devoted to the interpretation of
the obtained density as an effective primordial dipole density inside the forward
jet.

We have to compute the
``$\mbox{dipole}\!+\!\mbox{(virtual)gluon}
\rightarrow\bar q q\!+\!\mbox{gluon}$''
amplitude, where the initial-state dipole of radius ${\vec r}$ 
is part of an onium, and the final-state gluon has a transverse
momentum ${\vec q}$.
The modulus $q\!\equiv\! |{\vec q}|$ is larger than a given scale $\mu$. 
Having in mind the fact that in a physical process the initial-state
dipole will be part of a hadron instead of an onium, 
we will consider in the following
that $\mu\!\gg\! 1/r$ whenever needed for technical purpose.

\begin{figure}
\begin{center}
\mbox{\epsfig{file=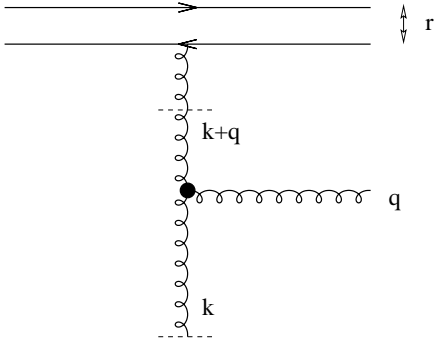,height=4.5cm}}
\end{center}
\caption{One of the diagrams contributing to the 
$\mbox{dipole}-g^*\rightarrow g\mbox{(forward)}+...$ cross-section.
{\it The horizontal dashed lines represent the points where $k_\bot$-factorization
is applied.}}
\label{fig:1}
\end{figure}

We start with the (virtual)gluon-dipole cross-section
$\hat\sigma_{g-d}$ which defines the gluon density inside a dipole at
lowest order in $\ab\!\equiv\!\alpha_sN_c/\pi$. 
It reads (see ref.\cite{Navelet:1997tx}):
\begin{equation}
f^0(\vec{k}^2)=\frac{\hat\sigma_{g-d}}{\vec{k}^2}=
\frac{\ab}{\vec{k}^2}\left(2\!-\!e^{i\vec{k}\cdot\vec{x}}
\!-\!e^{-i\vec{k}\cdot\vec{x}}\right)\ .
\label{eq:1}
\end{equation}
It can be expressed as an inverse Mellin-transform
in the transverse plane:
\begin{equation}
f^0(\vec{k}^2)=\frac{4\ab}{\vec{k}^2}
\int\frac{d\sigma}{2i\pi}\left(kr\right)^{2\sigma} v(\sigma)\ ,
\end{equation}
where $v$ is interpreted as the well-known dipole-gluon ``vertex'' in
the Mellin space:
\begin{equation}
v(\sigma)=\frac{2^{-2\sigma-1}}{\sigma}\frac{\G(1\!-\!\sigma)}
{\G(1\!+\!\sigma)}\ .
\label{eq:4}
\end{equation}
We then factorize the emission of a real gluon. It can be computed directly 
in the high-energy limit by evaluating the relevant graphs (one of
them is pictured in
fig.{\bf\ref{fig:1}}), but it is also convenient to see it as one step of the
BFKL ladder.
The initial gluon 
density $f^0$ and the density $f$ after emission of a gluon 
are related through the formula:
\begin{equation}
f(x,\vec{k}^2)={\ab}\left(\log\frac{1}{x}\right)
\int\frac{d^2\vec{q}}{\pi\vec{q}^2}
\theta(\vec{q}^2\!-\!\mu^2)
f^0(|\vec{k}\!+\!\vec{q}|^2)\ ,
\label{eq:mellin}
\end{equation}
which is the lowest order (in $\alpha_s$) BFKL equation written in an
unfolded form (see for instance ref.\cite{Kwiecinski:1995pu}).
The variable $x$ is proportional to $|t|/s$,
where $s$ and $t$ are the ordinary Mandelstam variables for the
reaction. In deep-inelastic
scattering, $x$ would stand for the Bjorken variable.

Let us write the Mellin-transform of eq.(\ref{eq:mellin}):
\begin{multline}
\frac{h(\gamma)}{\gamma}\equiv
\int_0^\infty\frac{d^2\vec{k}}{\pi\vec{k}^2}|\vec{k}|^{2\gamma}
f(x,\vec{k}^2)
=\ab\left(\log\frac{1}{x}\right)\times\\
\times\int\frac{d^2\vec{q}}{\pi\vec{q}^2}\theta(\vec{q}^2\!-\!\mu^2)
\int \frac{d^2\vec{k}}{\pi\vec{k}^2}
|\vec{k}|^{2\gamma} f^0(|\vec{k}\!+\!\vec{q}|^2)\ ,
\label{eq:hgsh}
\end{multline}
where we used similar notations as in ref.\cite{Catani:1991eg},
although we have not performed the Mellin-transform with respect to
$x$. 
The variable $\gamma$ describes a path 
$]\gamma_0\!-\!i\infty,\gamma_0\!+\!i\infty[$
in the complex plane, with $0\!<\!\real\,\gamma_0\!<\!1/2$.
Inserting eqs.(\ref{eq:1})-(\ref{eq:4}) into eq.(\ref{eq:hgsh}), it follows that:
\begin{multline}
\frac{h(\gamma)}{\gamma}=4\ab^2\left(\log\frac{1}{x}\right)
\int\frac{d\sigma}{2i\pi}v(\sigma)r^{2\sigma}\times\\
\times\int\frac{d^2\vec{q}}{\pi\vec{q}^2}\theta(\vec{q}^2\!-\!\mu^2)
\int\frac{d^2\vec{k}}{\pi}|\vec{k}|^{2\gamma-2}|\vec{k}\!-\!\vec{q}|^{2\sigma-2}\ .
\end{multline}
The integration over $\vec{k}$ can easily be performed switching to
complex variables and using the well-known identity 
(see \cite{Geronimo:2000fj} and references therein):
\begin{multline}
\int\frac{dz d\bar z}{2i}|z|^{2\alpha-2}|1-z|^{2\beta-2}\\
=\pi\,\frac{\G(\alpha)\G(\beta)}{\G(\alpha\!+\!\beta)}
\frac{\G(1\!-\!\alpha\!-\!\beta)}{\G(1\!-\!\alpha)\G(1\!-\!\beta)}\ ,
\end{multline}
valid for $\real\,\alpha$, $\real\,\beta>0$ and $\real(\alpha\!+\!\beta)<2$.
The result reads:
\begin{multline}
\frac{h(\gamma)}{\gamma}=2\ab^2\left(\log\frac{1}{x}\right)
\frac{\G(\gamma)}{\G(1\!-\!\gamma)}\mu^{2\gamma-2}\times\\
\times G^{40}_{35}\left(\left.
\matrix {1,1,2\!-\!\gamma}\\
{0,0,1\!-\!\gamma,1\!-\!\gamma,1\!-\!\gamma}
\endmatrix\right|
\left(\frac{\mu r}{2}\right)^2\right)\ ,
\end{multline}
where the Meijer-function $G^{40}_{35}$ (which arguments will be
abreviated in the following) writes:
\begin{equation}
G^{40}_{35}(\gamma,\mu r)\!=\!
\int_{\mathcal C}\frac{d\sigma}{2i\pi}\left(\frac{\mu r}{2}\right)^{2\sigma}
\frac{1}{\sigma^2(1\!-\!\gamma\!-\!\sigma)}
\frac{\G(1\!-\!\gamma\!-\!\sigma)}{\G(\gamma\!+\!\sigma)}\ ,
\end{equation}
the contour of integration ${\mathcal C}$ being defined on fig.{\bf 2}.
The function $h(\gamma)/\gamma$ can be seen as the coefficient-function 
in the sense of ref.\cite{Catani:1991eg}, for the onium+forward-jet system.
\begin{figure}
\begin{center}
\mbox{\epsfig{file=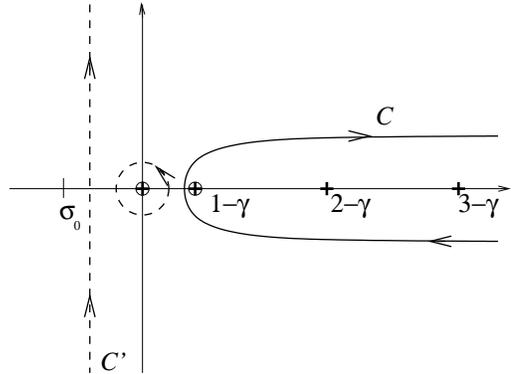,height=5cm}}
\end{center}
\caption{Contour of integration.
{\it Crosses: poles of the integrand. Circles: 
double poles. Dashed line: contour of integration after deformation.}}
\label{fig:2}
\end{figure}

\section{Interpretation in the colour-dipole model.}

In ref.\cite{Munier:1998vk}, a relation was established between the
coefficient-function $h(\gamma)/\gamma$ of a virtual photon and its
corresponding squared wave-function $\varphi(\gamma)$ on a dipole basis.
We shall make use of it in the present context of semi-exclusive
factorization to extract
the dipole content of the forward jet from the coefficient-function 
found above, at DLL accuracy. The double-logs should manifest
themselves through factors like $\ab\log(1/x)\log(r/r_0)$, where
$r_0$ is a characteristic size for the final state.
The relationship between the squared wave-function and the
coefficient-function reads:
\begin{equation}
\varphi(\gamma)=\frac{1}{\ab}\frac{h(\gamma)}{\gamma}\frac{1}{v(1\!-\!\gamma)}\ .
\end{equation}
An intuitive way of understanding this relation could be the following:
dividing the coefficient-function pictured in fig.{\bf\ref{fig:1}} 
by the factor $v(1\!-\!\gamma)$ amounts to getting rid of the
vertex of the lowest gluon which in this picture is also a
gluon-dipole vertex. Hence one ends up with the dipole
content of the scattering object:
\begin{equation}
\varphi(\gamma)=4\ab\left(\log\frac{1}{x}\right)
\left(\frac{\mu}{2}\right)^{2\gamma\!-\!2}
(1\!-\!\gamma)^2\times G^{40}_{35}(\gamma,\mu r)\ .
\end{equation}

Let us explore the limit $\mu r\!\gg\!1$, in which the forward jet has a
transverse momentum much larger than the characteristic scale of the
initial onium. The Meijer-function $G^{40}_{35}$ can be approximated in a
straightforward manner by picking the pole (at $\sigma\!=\!0$) which
lies on the left of the integration path. Indeed, the contour
${\mathcal C}$ can be deformed to ${\mathcal C}^\prime$ (see
fig.{\bf 2}) since the integral converges on any path such that
$\real\,\sigma\!>\!\sigma_0$, where $\sigma_0\!\equiv\!
-1/2\!-\!\real\,\gamma$:
\begin{align}
G^{40}_{35}(\gamma,\mu r)
&=\int_{\mathcal C}\frac{d\sigma}{2i\pi}
\left(\frac{\mu r}{2}\right)^{2\sigma}
\frac{1}{\sigma^2(1\!-\!\gamma\!-\!\sigma)}
\frac{\G(1\!-\!\gamma\!-\!\sigma)}{\G(\gamma\!+\!\sigma)}\nonumber\\
&=\left.\frac{\partial}{\partial\sigma}\right|_{\sigma=0}
\left(\left(\frac{\mu r}{2}\right)^{2\sigma}\frac{1}{1\!-\!\gamma\!-\!\sigma}
\frac{\G(1\!-\!\gamma\!-\!\sigma)}{\G(\gamma\!+\!\sigma)}\right)+\nonumber\\
&\
+\!\int_{\mathcal C^\prime}\frac{d\sigma}{2i\pi}
\left(\frac{\mu r}{2}\right)^{2\sigma}
\frac{1}{\sigma^2(1\!-\!\gamma\!-\!\sigma)}
\frac{\G(1\!-\!\gamma\!-\!\sigma)}{\G(\gamma\!+\!\sigma)}.
\end{align}
The integral taken on the contour ${\mathcal C}^\prime$ is subdominant
by some power of $2/(\mu r)$ with respect to the
contribution of the double-pole at $\sigma\!=\!0$, and one writes:
\begin{multline}
G^{40}_{35}(\gamma,\mu r)\!=\!\frac{\G(1\!-\!\gamma)}{(1\!-\!\gamma)\G(\gamma)}
\bigg\lbrace 2\log\frac{\mu
r}{2}\!-\!\psi(\gamma)\!-\!\psi(1\!-\!\gamma)\!+\\
+\!\frac{1}{1\!-\!\gamma}\bigg\rbrace\!+\!
\left\{\mbox{terms suppressed by powers of $1/(\mu r)$}\right\}\ .
\end{multline}
This approximation proves to be very good numerically even
for relatively small $\mu r$.

Then the squared dipole wave-function reads, in this approximation:
\begin{multline}
\varphi(\gamma)=4\ab\left(\log\frac{1}{x}\right)
\left(\frac{\mu}{2}\right)^{2\gamma-2}
\frac{\G(2\!-\!\gamma)}{\G(\gamma)}
\left(2\log\frac{\mu r}{2}\right.\\
\left.-\psi(\gamma)-\psi(1-\gamma)\!+
\!\frac{1}{1-\gamma}\right)\ .
\label{eq:phigam}
\end{multline}

We want to obtain an expression for the distribution of dipoles in the system
in coordinate space, i.e. as a function of the transverse
size $\rho$. It is given by the inverse-Mellin
transform of $\varphi(\gamma)$:
\begin{equation}
\varphi(\rho)=\frac{1}{\rho^2}\int\frac{d\gamma}{2i\pi}
\rho^{2\gamma-2}\varphi(\gamma)\ .
\end{equation}
We obtain the following result:
\begin{equation}
\varphi(\rho)=8\ab\left(\log\frac{1}{x}\right)
\left(\left(\log\frac{r}{\rho}\right)
\frac{\mu}{\rho}J_1(\mu\rho)+\frac{1}{\rho^2}J_0(\mu\rho)\right)\ .
\end{equation}
The second term is not relevant in our approximation, since
the limit of large $\mu r$ selects the DLL; the terms beyond 
this approximation are not under control. Hence the interaction of the
system formed by the initial dipole and the forward gluon can be
viewed as a dipole-dipole interaction provided the system is
described by the following dipole distribution:
\begin{equation}
\varphi_{{\mathrm{DLL}}}(\rho)=
8\ab\left(\log\frac{1}{x}\right)\left(\log\frac{r}{\rho}\right)
\frac{\mu}{\rho}J_1(\mu\rho)\ .
\label{eq:dipolegluon}
\end{equation}

Let us give an interpretation of the various factors in this
distribution. First, note that the dependence on the initial dipole size $r$ 
only appears in the factor $\log(r/\rho)$. This remarkable fact technically
results from the combination of the term $\log(\mu r/2)$  and the
inverse-Mellin transform of the $\psi$ functions in the inverse-Mellin
transform of eq.(\ref{eq:phigam}).
Physically, the overall
factor $2\ab\log(1/x)\log(r/\rho)$ can then be interpreted as the
probability of finding a dipole of size $\rho$ inside a dipole of size
$r$. Indeed, in the DLL approximation at lowest order in $\alpha_s$ and
assuming an available energy proportional to $1/x$, the
probability of finding a dipole of size $|\vec\varrho|$ between
$|\vec\rho|$ and $|\vec r|$ 
inside a dipole of size $\vec r$ reads \cite{Mueller:1994rr}:
\begin{multline}
\ab\left(\log\frac{1}{x}\right)
\int_{\rho^2}^{r^2}\frac{d^2\varrho}{\pi}\,\frac{{\vec r}^2}{{\vec\varrho}^2
({\vec r\!-\!\vec\varrho})^2}\\
\simeq
 \ab\left(\log\frac{1}{x}\right)\int_{\rho^2}^{r^2}
 \frac{d^2\varrho}{\pi{\vec\varrho}^2}
=2\ab\log\frac{1}{x}\log\frac{r}{\rho}\ ,
\end{multline}
where the second approximate equality holds when $\varrho\ll r$.
Dividing out this factor in eq.(\ref{eq:dipolegluon}) leads to 
the universal dipole content of the gluon radiated into the final
state. We normalize the first moment of
the obtained squared wave-function to unity\footnote{
This choice means that $\phi$ has now the unusual dimension
$[\mbox{mass}]^3$, but it enables a comparison to the photon 
squared wave-function $\phi^{\gamma^*}$, for which the $\gamma=0$ 
moment diverges logarithmically.} (i.e. $\phi(\gamma\!=\!1/2)\equiv
1$), and so we are led to:
\begin{equation}
\phi(\rho)=\frac{\mu^2}{\rho}J_1(\mu\rho)\ .
\end{equation}
Note that this squared wave-function for which we provide a derivation
here is exactly the one (integrated over the
energy-share variable $z$ and properly normalized) 
that was postulated in ref.\cite{Peschanski:1999hf}.
To see this, we only need to recall that the Mellin transform of the function
$J_1(\mu\rho)/\rho$ reads:
\begin{equation}
\int_0^\infty\frac{d^2\vec\rho}{\pi}|\vec\rho|^{2-2\gamma}
\frac{J_1(\mu\rho)}{\rho}=\mu^{2\gamma-3}2^{2-2\gamma}
\frac{\G(2\!-\!\gamma)}{\G(\gamma)}\ .
\end{equation}

The behaviour of $\phi$ is represented in fig.{\bf 3}. 
It is compared
to the behaviour of the transverse virtual photon squared
wave-function in $q\bar q$ pairs, 
integrated over the fraction of the photon longitudinal
momentum $z$ carried by the quark. This photon wave-function reads:
\begin{multline}
\phi^{\gamma^*}(\rho)=
\frac{64\mu}{9\pi^3}\int_0^1dz\,\mu^2z(1\!-\!z)(z^2\!+\!(1\!-\!z)^2)\times\\
\times K_1^2(\mu\rho\sqrt{z(1\!-\!z)})\ ,
\end{multline}
where the same normalization $\phi^{\gamma^*}(\gamma\!=\!1/2)\!\equiv\!1$ 
has been enforced.

\begin{figure}
\begin{center}
\mbox{\epsfig{file=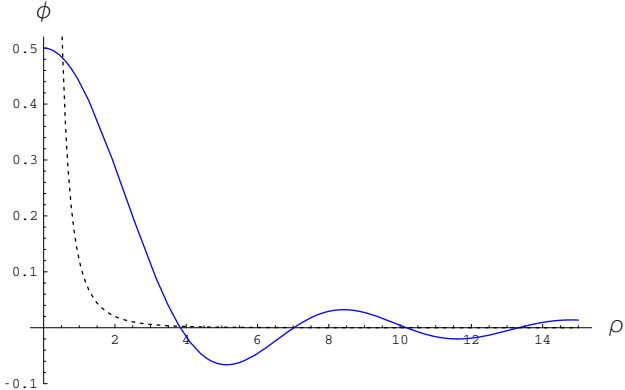,height=5.2cm}}
\end{center}
\caption{Dipole density as a function of the transverse size.
{\it The scale $\mu$ has been set to 1.
Continuous line: forward-jet. Dashed line: (transverse)
virtual photon.}}
\label{fig:3}
\end{figure}
Some remarks are in order.
The squared wave-function $\phi$
represents the effective distribution of
dipoles resulting from a final-state gluon which has its transverse momentum
larger than $\mu$. We note that it is slowly decreasing with $\rho$
($\sim \rho^{-3/2}$), which means that dipoles of large size occur
with a non-negligible probability. The width of the distribution
is of order $1/\mu$, but since its behaviour at infinity is only
powerlike, $\mu$ is not a clean cutoff.
It has an oscillatory behaviour, the oscillation length being
of the order of $1/\mu$. However, $\phi$ takes also negative
values, which indicates that it does not allow for a direct
probabilistic interpretation like in the case of the $q\bar{q}$-pair
distribution $\phi^{\gamma^*}$ inside the photon.

\section{Conclusions and outlook.}

Using a straightforward model and a QCD calculation in the
DLL approximation, we have shown that a gluonic hard probe 
in the final-state can be characterized by a dipole-distribution 
$\phi(\rho)$. This distribution decreases weakly with $\rho$ and 
takes negative values. We confirm the (integrated over $z$)
result obtained in ref.\cite{Peschanski:1999hf}.

The obtained distribution can now be used to compute processes
of several topologies: for instance deep-inelasic scattering on a
forward-jet at small-$x$, $p\!-\!\bar{p}$ interactions
with two jets in the final-state
separated by a large rapidity range.
In any case, the dipole formulation for these observables is of the
type:
\begin{multline}
{\mathcal O}\propto\int\frac{d\gamma}{2i\pi}\frac{\as^2}{\gamma^2(1\!-\!\gamma)^2}
\left(\frac{\mu}{Q_0}\right)^{2\gamma}
e^{\ab\chi(\gamma)Y}\times\\
\times\int d^2\rho\,\rho^{2\gamma}\phi(\rho)
\int d^2\rho_t\,\rho_t^{2-2\gamma}\phi_t(\rho_t)\ ,
\end{multline}
where
$\chi(\gamma)\!=\!2\psi(1)\!-\!\psi(\gamma)\!-\!\psi(1\!-\!\gamma)$
is the eigenvalue of the BFKL kernel, and $Y$ the rapidity range
between the two probes characterized by the dipole distributions $\phi$ 
and $\phi_t$. Note that the simplest observable one can compute using
this formula is the cross-section $\sigma$ for the production of
dijets of respective transverse momenta larger than $k_1$ and $k_2$
in hadronic collisions. It reads:
\begin{equation}
\sigma\propto\frac{\as^2}{k_1k_2}\int\frac{d\gamma}{2i\pi}\frac{1}{\gamma(1\!-\!\gamma)}
\left(\frac{k_1}{k_2}\right)^{2\gamma}
e^{\ab\chi(\gamma)Y}\ ,
\end{equation}
which is the Mueller-Navelet formula \cite{Mueller:1987ey}, modulo
some normalization we did not keep precise track of. Hence
this little calculation provides a further check of our dipole distribution.

This study may deserve several further investigations and
improvements. First of all, we worked
only up to terms beyond the DLL approximation.
It would be useful to perform a more complete calculation, by
computing to leading log-$1/x$ precision
all the graphs of the type of the one in fig.{\bf 1} which contribute,
to see if the dipole factorization still holds.

On the other hand, it would be nice to supplement our indirect method
of extracting the dipole content by a more
direct calculation in the framework of the colour-dipole model: this
would insure the full control of the leading log-$1/x$, including a
correct treatment of the virtual corrections. Although quite
straightforward to formulate, the latter calculation exhibits many
technical difficulties. Both these proposed improvements deserve more studies.\\

\noindent{\bf Acknowledgements:}\\

\noindent
I thank R. Peschanski and H. Navelet for many useful suggestions
and a careful reading of the manuscript.

%-------------------------------------------------------------------------------------

\end{document}